
\tolerance=10000
\magnification=\magstep1
\baselineskip=24pt
\vsize=7.5in
\hsize=5in
\hoffset=.25in
\ \vskip 1in
\centerline{\bf STUDY OF $\sigma(750)$ AND $\rho^0(770)$ PRODUCTION}
\centerline{\bf IN MEASUREMENTS OF $\pi N_\uparrow \to \pi^+ \pi^- N$ ON}
\centerline{\bf A POLARIZED TARGET AT 5.98, 11.85 AND 17.2 GeV/c.}
\bigskip
\bigskip
\centerline{M. Svec\footnote*{Electronic address:\ \
svec@hep.physics.mcgill.ca}}
\medskip
\centerline{Physics Department, Dawson College, Montreal, Quebec,
Canada H3Z 1A4}
\smallskip
\centerline{and}
\smallskip
\centerline{McGill University, Montreal, Quebec, Canada H3A 2T8}
\vfill\eject
\noindent
{\bf ABSTRACT}
\medskip
We present a new and improved model independent amplitude analysis
of reactions $\pi^+ n_\uparrow \to \pi^+ \pi^- p$ at 5.98 and 11.85
GeV/c and $\pi^- p_\uparrow \to \pi^- \pi^+ n$ at 17.2 GeV/c
measured with transversely polarized targets at the CERN Proton
Synchrotron. For dipion masses below 1000 MeV the pion production
process is described by two $S$-wave and six $P$-wave production
amplitudes. Previous analyses suffered from the presence of
unphysical solutions for moduli of amplitudes or cosines of their
relative phases, causing uncertainties regarding the signal for
scalar state $I=0\ 0^{++}(750)$. To remove the unphysical solutions
we use a Monte Carlo approach to amplitude analysis. In each $(m,t)$
bin we randomly varied the input spin density matrix elements 30
000 times within their experimental errors and performed amplitude
analysis for each selection. Unphysical solutions were rejected and
the physical solutions produced a continuous range of values for
moduli, cosines of relative phases and for partial wave intensities.
A clear signal for $\sigma(750)$ resonance emerges in all four
solutions for $S$-wave intensity $I_S$ at 5.98 and 11.85 GeV/c and
in both solutions for $S$-wave amplitude $|\overline S|^2\Sigma$ at
17.2 GeV/c. Its $\pi^+\pi^-$ decay width is estimated to be in the
range 200--300 MeV. We find significant suppression of $\rho^0$
production in the amplitudes $|U|^2\Sigma$, $|\overline N |^2\Sigma$
and, at 17.2 GeV/c, in $|L|^2\Sigma$. The mass dependence of
amplitudes $|\overline L |^2\Sigma$ and $|L|^2\Sigma$ shows
unexpected structures within the $\rho^0$ mass region which
correlate the mass spectra corresponding to opposite nucleon spins.
These features of $P$-wave moduli reveal the essential role of
nucleon spin in pion production process and contradict the
factorization hypothesis. Our results emphasize the need for a
systematic study of pion production on the level of amplitudes in a
new generation of dedicated experiments with spin at the recently
proposed high-intensity hadron facilities.
\vfill\eject
\noindent
{\bf 1. INTRODUCTION}
\medskip
The pion production process $\pi N\to \pi^+\pi^-N$ has always
served to develop our ideas on the dynamics of hadron collisions and
hadron production. In 1978, Lutz and Rybicki showed$^1$ that
measurements of pion production in meson-nucleon scattering on
transversely polarized target yield in a single experiment enough
observables that almost complete and model independent amplitude
analysis can be performed. In the kinematic region with dimeson
masses below 1000 MeV the dimeson system is produced predominantly
in spin states $J=0$ ($S$-wave) and $J=1$ ($P$-wave). The results
enable us to study the pion production and the resonance production
on the level of spin-dependent amplitudes rather than spin-averaged
cross-sections. In particular, a model independent separation of
$S$-wave and $P$-wave amplitudes is possible only in measurement on
transversely polarized targets.

The high statistics measurement of $\pi^- p \to \pi^-\pi^+ n$ at
17.2 GeV/c at CERN-PS on unpolarized target$^2$ was later repeated
with a transversely polarized target$^{\rm 3--7}$ at the same
energy. The resulting model independent amplitude analysis$^{3,4,5}$
provided the first evidence for significant contributions from
helicity-nonflip amplitudes with $A_1$ exchange quantum numbers
$(I^G=1^-, J^{PC}=1^{++})$ which had long been assumed absent.
Moreover, the $S$-wave partial wave intensity $I_S$ showed a clear
bump in the 750--800 MeV mass region$^8$ in one of the two
solutions$^{4,5,6}$ which indicated the possibility of a new scalar
state with a mass near $\rho^0$.

Additional information was provided by the first measurement of
$\pi^+ n \to \pi^+\pi^- p$ reaction on polarized deuteron target at
5.98 and 11.85 GeV/c also done at CERN-PS.$^{9,10,11}$ Amplitude
analysis at larger momentum transfer confirmed the evidence for
large $A_1$-exchange contributions and found resonant-like
structures in the $S$-wave partial-wave intensities.$^{12}$ This
data also found important $t$-dependent structures in the moduli of
$P$-wave amplitudes within the $\rho^0$ mass region.$^{13}$

In a recent paper$^{14}$ we focused on the evidence for the scalar
state $I=0$ $0^{++}(750)$ coming from all these measurements on
polarized targets. We found that all four solutions for $S$-wave
partial wave intensity at 5.98 and 11.85 GeV/c  suggest resonance
structure around 750 MeV. Its $\pi^+\pi^-$ decay width depends on
the solution and was estimated to be in the range of 100--250 MeV.
The evidence for this state is strengthened by the fact that the
$S$-wave amplitudes $S$ and $\overline S$ are nearly phase
degenerate with the dominant resonating $P$-wave amplitudes $L$ and
$\overline L$. It was suggested$^{14}$ that the $I=0\ 0^{++}(750)$
state is best understood as the lowest-mass scalar gluonium
$0^{++}(gg)$. Our results are in agreement with the $S$-wave
partial-wave intensity for $\pi^+ \pi^- \to \pi^0 \pi^0$
estimated$^{15}$ from the measurement of $\pi^+ p \to \pi^0
\pi^0\Delta^{++}$ at 8.0 GeV/c.

The data on transversely polarized targets are best analysed in
terms of nucleon transversity amplitudes. There are two $S$-wave and
six $P$-wave amplitudes. Amplitude analysis expresses
analytically$^{1,12}$ the eight moduli and six cosines of relative
phases of nucleon transversity amplitudes in terms of measured spin
density matrix (SDM) elements. There are two similar solutions.
However, in many $(m,t)$ bins the solutions are unphysical:
typically a cosine has magnitude larger than one or the two
solutions for moduli are complex conjugate with a small imaginary
part. In Ref.~12 and 14 we presented the direct analytical results
(taking only real part of complex solutions). The authors of Ref.
3--6 took these analytical solutions as starting values for a
$\chi^2$ minimization program which fitted the measured observables
to obtain physical values of moduli and cosines.

The occurrence of unphysical solutions is a major difficulty for
all amplitude analyses of pion production and the source of
uncertainty in the evidence for $I=0\  0^{++}(750)$ resonance. In
this paper we investigate this problem using Monte Carlo
methods.$^{16, 17}$

The basic idea of Monte Carlo amplitude analysis is to filter
out$^{16}$ the unwanted unphysical solutions and to determine$^{17}$
the range of physical values of moduli and cosines of relative
phases. To achieve this we randomly varied the input SDM elements
within their errors, performed the amplitude analysis for each new
set of the input SDM elements, and retained the resulting moduli and
cosines only when all of them had physical values in both
solutions. The results presented in this report are based on 30 000
random variations of input SDM elements in each $(m,t)$ bin. The
distributions of moduli and cosines define the range of their
physical values and their average  values$^{17}$ in each $(m,t)$
bin.

The results for the $P$-wave moduli are essentially the same as in
the previous analysis$^{14}$ but the changes for the $S$-wave are
striking. After filtering out the unwanted unphysical solutions, a
clear signal for $I=0\ 0^{++}(750)$ state emerges in all four
solutions for $S$-wave partial wave intensity at all 3 energies.

The paper is organized as follows. In Section 2 we review the basic
formalism. In Section 3 we describe our Monte Carlo approach to
amplitude analysis of $\pi N \to \pi^+ \pi^- N$ reactions on
polarized target. The evidence for the $I=0\  0^{++}(750)$ state is
presented in Section 4. In Section 5 we describe the spin dependence
of $\rho^0$ production and discuss its unexpected features. In
Section 6 we discuss the assumptions involved in the determination
of $\pi\pi$ phase shifts from data on $\pi^- p \to \pi^- \pi^+ n$ on
unpolarized target and present tests of the key assumption of
absence of $A_1$-exchange in measurements on polarized targets. The
paper closes with a summary in Section 7.
\bigskip
\noindent
{\bf 2. BASIC FORMALISM}
\medskip
For invariant masses below 1000 MeV, the dipion system in reactions
$\pi N \to \pi^+ \pi^- N$ is produced predominantly in spin states
$J=0$ ($S$-wave) and $J=1$ ($P$-wave). The experiments on
transversely polarized targets then yield 15 spin-density-matrix
(SDM) elements describing the dipion angular distribution. The
measured SDM elements are$^{10, 11}$
$$\rho_{ss} + \rho_{00} + 2\rho_{11},\ \rho_{00} - \rho_{11},\
\rho_{1-1}\eqno(2.1a)$$
$$Re \rho_{10},\ Re \rho_{1s},\ Re \rho_{0s}$$
$$\rho_{ss}^y + \rho_{00}^y + 2 \rho_{11}^y,\ \rho_{00}^y -
\rho_{11}^y,\ \rho_{1-1}^y\eqno(2.1b)$$
$$Re \rho_{10}^y,\ Re \rho_{1s}^y,\ Re \rho_{0s}^y$$
$$Im\ \rho_{1-1}^x,\ Im \rho_{10}^x,\ Im \rho_{1s}^x\eqno(2.1c)$$
\noindent
The SDM elements (2.1a) are also measured in experiments on
unpolarized targets. The observables (2.1b) and (2.1c) are
determined by the transverse component of target polarization
perpendicular and parallel to the scattering plane $\pi N \to (\pi^+
\pi^-)N$, respectively. The SDM elements (2.1) depend on $s,t,m$
where $s$ is the c.m. system energy squared, $t$ is the
four-momentum transfer squared, and $m$ is the $\pi^+\pi^-$
invariant mass. There are two linear relations among the matrix
elements in (2.1):
$$\rho_{ss} + \rho_{00} + 2 \rho_{11} = 1\eqno(2.2)$$
$$\rho_{ss}^y + \rho_{00}^y + 2 \rho_{11}^y = A$$
\noindent
where $A$ is the polarized target asymmetry.

The reaction $\pi^+ n \to \pi^+ \pi^- p$ is described by pion
production amplitudes $H_{\lambda_p, 0\lambda_n}$
$(s,t,m,\theta,\phi)$ where $\lambda_p$ and $\lambda_n$ are
helicities of the proton and neutron, respectively. The angles
$\theta, \phi$ describe the direction of $\pi^+$ in the $\pi^+
\pi^-$ rest frame. The production amplitudes can be expressed in
terms of production amplitudes corresponding to definite dipion spin
$J$ using an angular expansion
$$H_{\lambda_p, 0\lambda_n} = \sum\limits^\infty_{J=0}
\sum\limits^{+J}_{\lambda = - J} (2J + 1)^{{1\over 2}}
H^J_{\lambda\lambda_p, 0\lambda_n} (s,t,m) d^J_{\lambda 0} (\theta)
e^{i\lambda\phi}\eqno(2.3)$$
\noindent
where $J$ is the spin and $\lambda$ the helicity of the
$\pi^+\pi^-$ dipion system. Our amplitude analysis is carried out in
the $t$-channel helicity frame for the $\pi^+\pi^-$ dimeson system.
The helicities of the initial and final nucleons are always in the
$s$-channel helicity frame.

The ``partial-wave'' amplitudes $H^J_{\lambda\lambda_p,
0\lambda_m}$ can be expressed in terms of nucleon helicity
amplitudes with definite $t$-channel exchange naturality. In the
case when the $\pi^+\pi^-$ system is produced in the $S$- and
$P$-wave states we have
$$0^-{1\over 2}^+\to 0^+{1\over 2}^+:\ H^0_{0+,0+}=S_0\eqno(2.4a)$$
$$\qquad\qquad\qquad\qquad H^0_{0+,0-} =S_1$$
$$0^-{1\over 2}^+ \to 1^- {1\over 2}^+:\ H^1_{0+,0+}=L_0\eqno(2.4b)$$
$$\qquad\qquad\qquad\qquad H^1_{0+,0-} = L_1$$
$$\qquad\qquad\qquad\qquad H^1_{\pm 1+, 0+}= {{N_0 \pm
U_0}\over{\sqrt 2}}$$
$$\qquad\qquad\qquad\qquad H^1_{\pm 1+,0-} = {{N_1 \pm
U_1}\over{\sqrt 2}}$$
\noindent
In (2.4), $0^-$ stands for pion, ${1\over 2}^+$ for nucleon, $0^+$
for $J=0$ dipion state (S-wave), and $1^-$ for $J=1$ dipion state
(P-wave). At large $s$, the amplitudes $N_0$ and $N_1$ are both
dominated by natural $A_2$ exchange. The amplitudes $S_n, L_n, U_n,
n=0,1$ are dominated by unnatural exchanges: $A_1$ exchange for
$n=0$ and $\pi$ exchange for $n=1$. The index $n=|\lambda_n -
\lambda_p |$ is nucleon helicity flip.

The data on transversely polarized targets are best analysed in
terms of nucleon transversity amplitudes (NTA's).$^{1,11,12}$ In our
kinematic region we work with two $S$-wave and six $P$-wave NTA's
of definite naturality defined as follows$^{1,11,12}$
$$S=(S_0 + iS_1)/\sqrt 2\ ,\ \overline S= (S_0 - iS_1)/\sqrt
2\eqno(2.5)$$
$$L=(L_0 + iL_1)/\sqrt 2\ ,\ \overline L = (L_0 - iL_1)/\sqrt 2$$
$$U= (U_0 + iU_1)/\sqrt 2\ ,\ \overline U = (U_0 - iU_1)/\sqrt 2$$
$$N=(N_0 - iN_1)/\sqrt 2\ ,\ \overline N = (N_0 + iN_1)/\sqrt 2$$
\noindent
The amplitudes $S, L, U,N$ and $\overline S, \overline L, \overline
U, \overline N$ correspond to recoil nucleon transversity ``down''
and ``up'', respectively.$^{11, 12}$ The ``up'' direction is the
direction of normal to the scattering plane defined according to the
Basel convention by ${\vec p}_\pi \times {\vec p}_{\pi\pi}$ where
${\vec p}_\pi$ and ${\vec p}_{\pi\pi}$ are the incident pion and
dimeson momenta in the target nucleon rest frame. The $S$-wave
amplitudes $S, \overline S$ and $P$-wave amplitudes $L, \overline L$
have dimeson helicity $\lambda = 0$. The pairs of amplitudes $U,
\overline U$ and $N, \overline N$ are combinations of nucleon
helicity amplitudes with dimeson helicities $\lambda = \pm 1$ and
have opposite $t$-channel-exchange naturality.

We can now express the observables in terms of amplitudes. In our
normalization, the integrated cross section $\Sigma \equiv
d^2\sigma/dmdt$ is given by
$$\eqalignno{\Sigma&=\sum\limits_{n=0,1} |S_n|^2 + |L_n|^2 +
|U_n|^2 + |N_n|^2&(2.6)\cr
&= |S|^2 + |\overline S |^2 + |L|^2 + |\overline L |^2 + |U|^2 +
|\overline U |^2 + |N|^2 + |\overline N |^2\cr}$$
\noindent
The cross section has not been measured in the experiments on
polarized targets. Consequently, we will work with normalized
amplitudes corresponding to
$$\Sigma = {{d^2\sigma}\over{dmdt}}\equiv1\eqno(2.7)$$
\noindent
Using (2.6), the relations for SDM elements in terms of normalized
helicity amplitudes read as follows.$^{1,12}$

Unpolarized SDM elements
$$\eqalignno{\rho_{ss} + \rho_{00} + 2\rho_{11}&= \sum\limits_{n=0}
|S_n|^2 + |L_n|^2 + |U_n|^2 + |N_n|^2&(2.8a)\cr
\rho_{00} - \rho_{11}&= \sum\limits_{n=0,1} |L_n|^2 - {1\over 2}
(|N_n|^2 + |U_n|^2)\cr
\rho_{1-1}&= \sum\limits_{n=0,1}{1\over 2} (|N_n|^2 - |U_n|^2)\cr
\sqrt 2 Re \rho_{10}&= \sum\limits_{n=0,1} Re (U_n L_n^*)\cr
\sqrt 2 Re \rho_{1s}&= \sum\limits_{n=0,1} Re (U_n S_n^*)\cr
Re \rho_{0s}&= \sum\limits_{n=0,1} Re (L_n S_n^*)\cr}$$
\noindent
Polarized SDM elements
$$\eqalignno{\rho_{ss}^y + \rho_{00}^y + 2\rho_{11}^y&= 2 Im (S_0
S_1^* + L_0 L_1^* + U_0 U_1^* + N_0 N_1^*)&(2.8b)\cr
\rho_{00}^y - \rho_{11}^y&= Im (2L_0 L_1^* - N_0 N_1^* - U_0 U_1^*)\cr
\rho_{1-1}^y&= Im (N_0 N_1^* - U_0 U_1^*)\cr
\sqrt 2 Re \rho_{10}^y&= Im (U_0 L_1^* - U_1 L_0^*)\cr
\sqrt 2 Re \rho_{1s}^y&= Im (U_0 S_1^*  - U_1 S_0^*)\cr
Re \rho_{0s}^y&= Im (L_0 S_1^* - L_1 S_0^*)\cr}$$
\bigskip
$$\eqalignno{- Im \rho_{1-1}^x&= Im (N_0 U_1^* + N_1 U_0^*)&(2.8c)\cr
\sqrt 2 Im \rho_{10}^x&= Im (N_0 L_1^* + N_1 L_0^*)\cr
\sqrt 2 Im \rho_{1s}^x&= Im (N_0 S_1^* + N_1 S_0^*)\cr}$$
\noindent
Only the polarization dependent SDM elements measure the nucleon
helicity flip-nonflip interference. The observables (2.8b) and
(2.8c) measure the interference between the amplitudes of the same
and opposite naturalities, respectively.

To express the observables in terms of normalized nucleon
transversity amplitudes (2.5), we first introduce partial wave
cross-sections $\sigma(A)$ and partial-wave
polarizations $\tau(A)$ defined for amplitudes $A = S, L, U, N$ as
$$\eqalignno{\sigma(A)&= |A_0|^2 + |A_1|^2 = |A|^2 + |\overline A
|^2&(2.9)\cr
\tau(A)&=2\epsilon Im (A_0 A_1^*) = |A|^2 - |\overline A |^2\cr}$$
\noindent
where $\epsilon = +1$ for $A=S,L,U$ and $\epsilon = -1$ for $A=N$.
In our normalization the reaction cross-section is
$$\Sigma = \sigma(S) + \sigma(L) + \sigma(U) + \sigma(N) = 1\eqno(2.10)$$
\noindent
The relations for SDM elements (2.8a) and (2.8b) in terms of
normalized nucleon transversity amplitudes (2.5) and quantities
(2.9) read
$$\eqalignno{
\rho_{ss} + \rho_{00} + 2\rho_{11}&= \sigma(S) + \sigma(L) +
\sigma(U) + \sigma(N)&(2.11a)\cr
\rho_{00} - \rho_{11}&= \sigma(L) - {1\over 2} [\sigma(U) + \sigma(N)]\cr
\rho_{1-1} &= - {1\over 2} [\sigma(U) - \sigma(N)]\cr
\rho_{ss}^y + \rho_{00}^y + 2\rho_{11}^y&= \tau(S) + \tau(L) +
\tau(U) - \tau(N)&(2.11b)\cr
\rho_{00}^y - \rho_{11}^y&= \tau(L) - {1\over 2} [\tau(U) - \tau(N)]\cr
\rho_{1-1}^y&= -{1\over 2} [\tau(U) + \tau(N)]\cr
\sqrt 2 Re \rho_{10}&= Re(UL^* + \overline U\  \overline L^*)&(2.12a)\cr
\sqrt 2 Re \rho_{1s}&= Re (US^* + \overline U\  \overline L^*)\cr
Re \rho_{0s} &= Re (LS^* + \overline L\  \overline S^*)\cr
\sqrt 2 Re \rho_{10}^y&= Re (UL^* - \overline U\  \overline
L^*)&(2.12b)\cr
\sqrt 2 Re \rho_{1s}^y&= Re (US^* - \overline U\  \overline S^*)\cr
Re \rho_{0s}^y&= Re (LS^* - \overline L\  \overline S^*)\cr}$$
\noindent
The relations (2.11) and (2.12) suggest to introduce new
observables which are the sum and difference of the SDM elements
(2.8a) and (2.8b). Using the notation of (2.2), the first group of
new observables reads
$$\eqalignno{
a_1&= {1\over 2} [1+A] = |S|^2 + |L|^2 + |U|^2 + |\overline N
|^2&(2.13a)\cr
a_2&= [(\rho_{00} - \rho_{11}) + (\rho_{00}^y - \rho_{11}^y)] =
2|L|^2 - |U|^2 - |\overline N |^2\cr
a_3&= [\rho_{1-1} + \rho_{1-1}^y] = |\overline N |^2 - |U|^2\cr
a_4&= {1\over\sqrt 2} [Re \rho_{10} + Re \rho_{10}^y] = |U||L|\cos
(\gamma_{\lower.5ex\hbox{$\scriptstyle {LU}$}})&(2.13b)\cr
a_5&= {1\over\sqrt 2} [Re \rho_{1s} + Re \rho_{1s}^y] = |U||S|\cos
(\gamma_{\lower.5ex\hbox{$\scriptstyle {SU}$}})\cr
a_6&= {1\over 2} [Re \rho_{0s} + Re \rho_{0s}^y] = |L||S|\cos
(\gamma_{\lower.5ex\hbox{$\scriptstyle {SL}$}})\cr}$$
\noindent
Similar equations relate the diffeence of SDM elements to
amplitudes of opposite transversity. The second group of observables
is defined as
$$\eqalignno{
\overline a_1&= {1\over 2} [1-A] = |\overline S |^2 + |\overline L
|^2 + |\overline U|^2 + |N|^2&(2.14a)\cr
\overline a_2&= [(\rho_{00} - \rho_{11}) - (\rho_{00}^y -
\rho_{11}^y)] = 2|\overline L |^2 - |\overline U |^2 - |N|^2\cr
\overline a_3&= [\rho_{1-1} - \rho_{1-1}^y] = |N|^2 - |\overline U |^2\cr
\overline a_4&= {1\over\sqrt 2} [Re \rho_{10} - Re \rho_{10}^y] =
|\overline U ||\overline L |\cos
(\overline\gamma_{\lower.5ex\hbox{$\scriptstyle {LU}$}})&(2.14b)\cr
\overline a_5&= {1\over\sqrt 2} [Re \rho_{1s} - Re \rho_{1s}^y] =
|\overline U ||\overline S |\cos
(\overline\gamma_{\lower.5ex\hbox{$\scriptstyle {SU}$}})\cr
\overline a_6&= {1\over 2} [Re \rho_{0s} - Re \rho_{0s}^y] =
|\overline L ||\overline S |\cos
(\overline\gamma_{\lower.5ex\hbox{$\scriptstyle {SL}$}})\cr}$$
\noindent
In the Equation (2.13b) and (2.14b) we have introduced explicitly
the cosines of relative phases between the nucleon transversity
amplitudes.

The SDM elements (2.8c) form the third group of
observables$^{1,12}$ which is not used in the present amplitude
analysis.

The first group (2.13) involves four moduli $|S|^2, |L|^2, |U|^2$
and $|\overline N |^2$, and three cosines of relative phases
$\cos(\gamma_{\lower.5ex\hbox{$\scriptstyle {SL}$}})$,
$\cos(\gamma_{\lower.5ex\hbox{$\scriptstyle {SU}$}})$ and
$\cos(\gamma_{\lower.5ex\hbox{$\scriptstyle {LU}$}})$. The second
group (2.14) involves the same amplitudes but with opposite nucleon
transversity. In Ref.~12 we derived analytical solution for these
amplitudes in terms of observables. For the first group we obtained
a cubic equation for $|L|^2 \equiv x$
$$ax^3 + bx^2 + cx + d=0\eqno(2.15)$$
\noindent
with coefficients $a,b,c,d$ expressed in terms of observables $a_i,
i=1,2,\ldots ,6$ (see Ref.~1, 12). The remaining moduli and the
cosines are given by expressions
$$\eqalignno{
|S|^2&= (a_1 + a_2) - 3|L|^2&(2.16)\cr
|U|^2&= |L|^2 - {1\over 2} (a_2 + a_3)\cr
|\overline N|^2&= |L|^2 - {1\over 2} (a_2 - a_3)\cr
\cos(\gamma_{\lower.5ex\hbox{$\scriptstyle {LU}$}})&=
{a_4\over{|L||U|}}\cr
\cos(\gamma_{\lower.5ex\hbox{$\scriptstyle {SU}$}})&=
{a_5\over{|S||U|}},\  \cos(\gamma_{\lower.5ex\hbox{$\scriptstyle
{SL}$}})= {a_6\over{|S||L|}}\cr}$$
\noindent
The solution for the second group (2.14) is similar.

The physical solutions of cubic equation (2.15) for $|L|^2$ must
produce physical and normalized moduli and physical values for the
cosines
$$0\le |A|^2 \le 1,\ A=L,S,U,\overline N\eqno(2.17)$$
$$-1 \le \cos \gamma_k \le 1,\ k=LU, SU, SL$$
\noindent
There are similar constraints on the solutions for $|\overline L
|^2$ in the second group.

The analytical solution of the cubic equation (2.15) is given in
Table 1 of Ref.~12. One solution of (2.15) is always negative and it
is rejected. The other two solutions are generally positive and
close. However in a number of $(m,t)$ bins we get unphysical values
for some cosines and in some cases also negative moduli of
amplitudes. In some $(m,t)$ bins the mean values of input SDM
elements yield complex solutions for $|L|^2$ or $|\overline L|^2$ or
both (with positive real parts). To filter out the unwanted
unphysical solutions we now turn to Monte Carlo amplitude analysis.
\bigskip
\noindent
{\bf 3. MONTE CARLO AMPLITUDE ANALYSIS}
\medskip
The origin of the presented Monte Carlo method lies in the problem
of error combination and propagation when the function of input
uncertain variables is highly nonlinear.$^{17}$ This is our case.
The analytical solutions for the amplitude $x=|L|^2$ from the cubic
equation (2.15) are highly nonlinear functions of the input
observables -- see Table 1 of Ref.~12. Even if the input spin
density matrix (SDM) elements have Gaussian distributions, the
solutions of the cubic equation (2.15) and the amplitudes (2.16) are
non-Gaussian distributions as the result of the nonlinearity. The
question arises how to estimate the errors on the amplitudes.

In general, when the errors on the input variables are small, one
can use a linear approximation to error propagation (Ref.~17). This
method yields symmetric errors using $1\sigma$ errors as input. This
approximation was used in our previous analyses. Since the errors
on polarized SDM elements are not actually small, this approximation
is not satisfactory.

In his review paper (Ref.~17) F.~James advocates the use of Monte
Carlo method as perhaps the only way to calculate the errors in case
of nonlinear functions which produce non-Gaussian distributions.
The method has the added advantage that it can separate the physical
and unphysical solutions, something the linear approximation and
the $\chi^2$ method cannot do. The Monte Carlo method was first used
to calculate errors of analytical solutions of a cubic equation by
H.~Palka$^{18}$ in an amplitude analysis of reactions $\pi^- p \to
K^+ K^- n$ and $\pi^- p \to K^0_s K^0_s n$ at 63 GeV/c.

In our Monte Carlo amplitude analysis the input SDM elements were
randomly varied within their experimental errors in each $(m,t)$ bin
and amplitudes were calculated for each new selection of SDM
elements. The input SDM elements were varied independently using a
multidimensional random number generator SURAND$^{19}$ with an
initial seed number 80629.0. Since any sequence of random numbers
generated by SURAND is reproducible, our results are also
reproducible.$^{20}$

Each selection of SDM elements yields two solutions for amplitudes
of group 1 (eqs. (2.13))
$$|S|^2,\ |L|^2,\ |U|^2,\ |\overline N|^2\eqno(3.1)$$
$$\cos (\gamma_{\lower.5ex\hbox{$\scriptstyle {LU}$}}),\
\cos(\gamma_{\lower.5ex\hbox{$\scriptstyle {SU}$}}),\
\cos(\gamma_{\lower.5ex\hbox{$\scriptstyle {SL}$}})$$
\noindent
and two solutions for amplitudes of group 2 (eqs. (2.14)
$$|\overline S|^2,\ |\overline L|^2,\ |\overline U|^2,\ |N|^2\eqno(3.2)$$
$$\cos (\overline\gamma_{\lower.5ex\hbox{$\scriptstyle {LU}$}}),\
\cos(\overline\gamma_{\lower.5ex\hbox{$\scriptstyle {SU}$}}),
\cos(\overline\gamma_{\lower.5ex\hbox{$\scriptstyle {SL}$}})$$
\noindent
In each group the solution was classified as physical only when all
4 moduli and all 3 cosines of relative phases had physical values.
The selection of SDM elements was classified as pass only when all
solutions for amplitudes (3.1) and (3.2) were physical solutions.
Otherwise the selection was classified as fail.

The Monte Carlo amplitude analysis program was run with a total
number of SDM elements selections $N_{tot} =$ 10~000, 20~000 and
30~000 in each $(m,t)$ bin. Each selection is classified as pass or
fail according to the above criteria. The passing rate, or the ratio
$N_{pass}/N_{tot}$, was nearly constant in each $(m,t)$ bin as
$N_{tot}$ was increased from 10~000 to 20~000 and 30~000 selections.
Consequently no further increases in $N_{tot}$ were attempted.
However, there are considerable variations of $N_{pass}$ from bin to
bin. In Figure 1 we show  the $m$-dependence of $N_{pass}$ for the
runs with $N_{tot}=30~000$ for reactions $\pi^- p_\uparrow \to \pi^-
\pi^+ n$ at 17.2 GeV/c and $\pi^+ n_\uparrow \to \pi^+\pi^- p$ at
5.98 GeV/c. The results at 11.85 GeV/c are similar. For some reason
$N_{pass}$ is lowest for $m\sim 800$ MeV at all 3 energies. At each
energy there is one bin which produced no physical solution
(880--900 MeV bin for 17.2 GeV/c, 360--440 MeV bin at 5.98 GeV/c,
and 820--870 MeV bin at 11.85 GeV/c). At 17.2 GeV we shall omit also
the mass bin 640--660 MeV where we found $N_{pass}=3$ only.

The pass and fail SDM elements are retained in each $(m,t)$ bin to
determine their distribution, range of values and average value. A
priori, the ranges and average values are expected to be different
for pass and fail selections and could depend on SDM elements.
However, we found that for all unpolarized SDM elements in all
$(m,t)$ bins the ranges of values for pass and fail Monte Carlo
selections are the same and coincide with the ranges given by the
original errors. Also surprisingly, the average values of pass and
fail unpolarized SDM elements are equal to the mean values of the
input SDM elements. For polarized SDM elements we reach the same
conclusion for fail Monte Carlo selections. For pass Monte Carlo
selections, the polarized SDM elements have smaller ranges in a few
$(m,t)$ bins with lowest $N_{pass}$ values and their average values
differ up to 6\% from the mean values of input SDM elements. The
situation for $\rho_{ss}^y + \rho_{00}^y + 2\rho_{11}^y$ is typical
and is illustrated in the Table 1 for $\pi^- p \to \pi^- \pi^+ n$ at
17.2 GeV/c.

The values of moduli and cosines obtained for each pass Monte Carlo
selection were collected to determine their range and average
values in each $(m,t)$ bin. To study the distribution of values of
moduli and cosines for all $N_{pass}$ selections, the moduli were
binned in bins of size 0.02 and the cosines in bins of size 0.04. An
example of such distributions is shown in Figure 2 where we present
solutions 1 and 2 for $\cos(\gamma_{\lower.5ex\hbox{$\scriptstyle
{SL}$}})$ at 5.98 GeV/c for $m=$520--600 MeV and $-t=$0.2--0.4
(GeV/c).$^2$ Notice the typical differences between the
distributions for Solutions 1 and 2. The distribution for solution 1
yields a range of $\cos \gamma_{\lower.5ex\hbox{$\scriptstyle
{SL}$}}$ from 0.52 to 1.00 with an average value 0.94. The
distribution for Solution 2 yields a range of $\cos
\gamma_{\lower.5ex\hbox{$\scriptstyle {SL}$}}$ from 0.31 to 1.00
with an average of 0.64. The differences between the two
distributions suggest that the $\chi^2$ mimimization approach used
in Ref. 3,4,5 to obtain physical solutions may not be dependable.

The resulting ranges of values and average values for all
unnormalized moduli and for cosines are shown in Figure 3 for $\pi^-
p\to \pi^-\pi^+ n$ at 17.2 GeV/c and in Figure 4 for $\pi^+ n \to
\pi^+ \pi^- p$ at 5.98 GeV/c. The figures show the mass dependence
of the two solutions for moduli and cosines for $-t =$0.005--0.2
(GeV/c)$^2$ at 17.2 GeV/c and $-t=$0.2--0.4 (GeV/c)$^2$ at 5.98
GeV/c. The unnormalized moduli $|\overline A|^2\Sigma$ and
$|A|^2\Sigma, A=S,L,U.N$ are calculated using
$\Sigma=d^2\sigma/dmdt$ from Ref. 2 at 17.2 GeV/c and from Ref. 10
at 5.98 and 11.85 GeV/c. A comparison with Figures 1 and 2 of
Ref.~14 reveals that the physical solutions for $S$-wave moduli and
for cosines are much smoother and continuous. The jitter which
characterized the unphysical solutions is removed by filtering out
the unphysical solutions. We discuss the features of the amplitudes
in detail in the next two sections.

For each pass Monte Carlo selection we also calculated the
normalized partial-wave cross-sections $\sigma_A$ and partial-wave
polarizations $\tau_A$ defined in (2.9) in order to obtain their
range of values and average values. Using experimental results for
$\Sigma=d^2\sigma/dmdt$, we then calculated partial wave intensities
$I_A = \sigma_A \Sigma$ where $A=S,L,U,N$. The solutions for
amplitudes with opposite transversities are entirely independent. We
can denote the two solutions for normalized nucleon transversity
amplitudes as $A(i)$ and $\overline A(j)$ with $i=1,2$ and $j=1,2$.
Because the moduli squared in (2.9) are independent, there is a
fourfold ambiguity in the partial-wave intensities. Using the
indices $i$ and $j$ to label the four solutions, we get
$$I_A(i,j) = [|A(i)|^2 + |\overline A(j)|^2]\Sigma\eqno(3.3)$$
\noindent
where $A=S,L,U,N$. The partial-wave intensities (3.3) are defined
in the physical region and depend on the dipion mass $m$ and
momentum transfer $t$.

The Monte Carlo amplitude analysis presented in this paper involves
one important simplification, namely the experimental data points
are represented by a rectangular distribution of width $2\sigma$.
The $2\sigma$ width is motivated by the requirement to study the
propagation of $1\sigma$ errors on experimental data through the
highly nonlinear expressions for the amplitudes in order to define
the errors on amplitudes. The uniformity of the experimental data
distributions is due to the use of the uniform random number
generator.

The actual experimental data are presumed to have Gaussian
distributions. To examine the effect of Gaussian distribution of
data on the results of amplitude analysis, we analysed the data at
17.2 GeV/c and 5.98 GeV/c, using an IBM Gaussian random number
generator SNRAND.$^{19}$

In this calculation the ``fail'' distributions of SDM elements are
approximately Gaussian with a broad range of half-size of
4.--4.5$\sigma$. The ``pass'' distributions of SDM elements (giving
the physical solutions for amplitudes) have a narrower range of
approximate half-size $2\sigma$. Most of the points lie within
$1\sigma$ half-interval. Between $1\sigma$ and $2\sigma$ the
``pass'' distributions have only low tails. This means that most
physical solutions lie within $1\sigma$ vicinity of central points
and none beyond $2\sigma$.

As expected, the distributions for amplitudes (moduli and cosines)
are non-Gaussian. Importantly, the average values of distributions
are essentially identical to the average values obtained previously.
The agreement is such that the black points in the figures do not
change. The range of distributions is now broader. The minimum and
maximum values change typically by 0--20\%. However this increase is
due to only low tails of distributions. The broadening of range of
values now makes more difficult the estimation of errors. The usual
calculation of standard deviation does not make sense in the case of
such highly non-Gaussian distribution as is e.g. Solution 1 for
$\cos\gamma_{SL}$ (Fig. 2).

Because the average values of amplitudes remain virtually the same
and the distributions for amplitudes acquire only low tails, there
is no significant change from the results presented in this paper,
and the simplification used in our Monte Carlo amplitude analysis is
an acceptable approximation.
\bigskip
\noindent
{\bf 4. EVIDENCE FOR A SCALAR STATE $I=0\  0^{++}(750)$}
\medskip
After filtering out the unphysical solutions, the evidence for the
scalar state $I=0\  0^{++}(750)$ becomes much more convincing than
were the previous indications.$^{14}$ The evidence is based on three
observations:
\item{(a)} Resonance structure of the $S$-wave partial wave
intensity $I_S$.
\item{(b)} The resonant structure of unnormalized $S$-wave moduli.
\item{(c)} The constant relative phase between the $S$-wave
amplitudes $S,\overline S$ and the dominant $P$-wave amplitudes
$L,\overline L$.
\medskip
We first look at the $S$-wave moduli at 17.2 GeV/c at lower
momentum transfers which are summarized in Fig.~5. We see that the
amplitude $|\overline S|^2\Sigma$ has a clear resonant structure in
both solutions. This is a change from previous results in Ref.~14
where only the solution 1 was clearly resonating. In both solutions
the maximum is at 750--770 MeV with a width at half-height estimated
at 175--200 MeV. The moduli $|S|^2\Sigma$ of opposite transversity
have smaller magnitudes and do not show a resonant structure. At
first sight this might be surprising, but we see the same effect
also in $\rho^0$ production amplitudes. For instance, while the
amplitudes $|\overline L|^2\Sigma$, $|\overline U|^2\Sigma$ and
$|N|^2\Sigma$ show resonant structures, the amplitudes
$|L|^2\Sigma$, $|U|^2 \Sigma$ and $|\overline N|^2\Sigma$ are
smaller and show less resonant structure. We can ascribe this effect
to the spin dependent dynamics of resonance production. The
difference of magnitudes of $|\overline S|^2\Sigma$ and
$|S|^2\Sigma$ is due to the presence of $A_1$-exchange i.e.
nonvanishing amplitude $S_0$.

The usual signature of a production resonance is a peak or bump in
the production cross-section. The $S$-wave partial-wave intensity
$I_S = (|S|^2 + |\overline S|^2)\Sigma$ at 17.2 GeV/c is shown in
Fig.~6. We find a clear resonant signal at or near 750 MeV in two
solutions $I_S(1,1)$ and $I_S(2,1)$. The other two solutions
$I_S(1,2)$ and $I_S(2,2)$ do not show a clear resonant structure. In
these solutions $I_S$ is rising up to 770 MeV, then it drops but
remains high. The cause of this behaviour is the nonresonating
amplitude $|S|^2\Sigma$ which is large above 800 MeV. We are thus
faced with an apparent ambiguity in the $S$-wave intensity observed
also in the earlier studies.$^{4,5,6}$ However more fundamental
evidence for $\sigma(750)$ production at 17.2 GeV/c at lower $t$ is
the fact that both solutions for the amplitude $|\overline
S|^2\Sigma$ resonate. We see the ambiguity in $I_S$ only because the
amplitudes $|\overline S|^2\Sigma$ and $|S|^2\Sigma$ happen to have
magnitudes which are not very different. We do not see any
ambiguity e.g. in the intensity $I_L$ because the amplitude
$|\overline L|^2\Sigma$ is much larger than $|L|^2\Sigma$ which also
remains high above 800 MeV.

As seen in Figure 3, the amplitudes $|\overline L|^2\Sigma$ and
$|L|^2\Sigma$ are dominant $\rho^0$ production amplitudes. The
relative phase between the $S$-wave amplitudes $\overline S$ and $S$
and the $P$-wave amplitudes $\overline L$ and $L$ is another
important piece of evidence for existence of the $I=0\  0^{++}(750)$
resonance. In Solution 1 both relative phases
$\overline\gamma_{\lower.5ex\hbox{$\scriptstyle {SL}$}}$ and
$\gamma_{\lower.5ex\hbox{$\scriptstyle {SL}$}}$ are consistent with
zero and we have a phase degeneracy of amplitudes $\overline S$ and
$\overline L$, and $S$ and $L$. In Solution 2 the relative phases
are not zero but are small and constant over the considered mass
region. Since the amplitude $\overline L$ is resonating, the phase
degeneracy with amplitude $\overline S$ suggests that $\overline S$
also resonates with a resonance mass near $\rho_0$.

The $S$-wave intensity at larger momentum transfers $-t = 0.2 -
0.4$ (GeV/c)$^2$ at 5.98 and 11.85 GeV/c is shown in Fig.~7 and 8,
respectively. There is no ambiguity at larger momentum transfers. At
both energies we find a clear resonant signal at or near 750 MeV in
all solutions. At 5.98 GeV/c, the width at half-height is estimated
to be 270 MeV. We note that at 11.85 GeV/c, the solutions
$I_S(2,1)$ and $I_S(2,2)$ are consistent with a narrow width of 150
MeV.

The resonant structure of moduli at larger momentum transfers
$-t=$0.2--0.4 (GeV/c)$^2$ at 5.98 GeV/c and 11.85 GeV/c is less
clear. At both energies it is the Solution 2 which has a stronger
indication for a resonance in both moduli $|\overline S|^2\Sigma$
and $|S|^2\Sigma$. Because of lower statistics, the errors on input
SDM elements are larger which results in larger ranges of values for
cosines of relative phases. Nevertheless, we also observe that the
relative phases $\overline\gamma_{\lower.5ex\hbox{$\scriptstyle
{SL}$}}$ and $\gamma_{\lower.5ex\hbox{$\scriptstyle {SL}$}}$ are
near zero in Solution 1, and are only slowly varying over the
critical $\rho^0$ mass region in Solution 2. The unusually larger
variation of $\cos(\overline\gamma_{\lower.5ex\hbox{$\scriptstyle
{SL}$}}), \cos(\gamma_{\lower.5ex\hbox{$\scriptstyle {SL}$}}),
\cos(\overline\gamma_{\lower.5ex\hbox{$\scriptstyle {SU}$}})$ and
$\cos(\gamma_{\lower.5ex\hbox{$\scriptstyle {SU}$}})$ at $m=$980 MeV
could be due to the presence of scalar resonance $f_0 (975)$.

We have also fitted the $S$-wave partial-wave intensities $I_S$ to
a Breit-Wigner form using the CERN optimization program
FUMILI.$^{21}$ The parametrization used had the general form
$$I_A (m) = N_A|BW_R|^2\eqno(4.1)$$
\noindent
where $A= S, L, U, N$ and
$$BW_R = ({m\over{\sqrt q}}) \sqrt{2J + 1} {{m_R\Gamma}\over{m^2_R
- m^2 - im_R \Gamma}}\eqno(4.2)$$
\noindent
In (4.1) $N_A$ is a normalization constant which includes square of
elasticity $x$ and isospin factor. In (4.2) $q$ is the $\pi^-$
momentum in the $\pi^-\pi^+$ rest frame
$$q=|{\vec q}\,| = \sqrt{0.25m^2 - \mu^2}\eqno(4.3)$$
\noindent
where $\mu$ is the pion mass. In (4.2), $J$, $m_R$ and $\Gamma$ are
spin, mass and width of the resonance. The mass dependent width is
$$\Gamma = \Gamma_R ({q\over{q_{\lower.5ex\hbox{$\scriptstyle
R$}}}})^{2J+1} {{D_J (q_{\lower.5ex\hbox{$\scriptstyle R$}}
r)}\over{D_J(q r)}}\eqno(4.4)$$
\noindent
where $q_R = q (m = m_R)$ and $D_J$ are centrifugal barrier
functions of Blatt-Weiskopf$^{22}$
$$D_0(qr)=1.\eqno(4.5)$$
$$D_1(qr) = 1. + (qr)^2$$
\noindent
In (4.5) $r$ is the interaction radius. The factor $m/\sqrt q$ in
(4.2) arises from the Chew-Low formula.

For $S$-wave $A=S$, $J=0$, $R=\sigma$ and $\Gamma$ is independent
of $r$. The results of the fit to the two resonating solutions for
$I_S$ at 17.2 GeV/c are presented in Table 2. We encountered
difficulties in fitting the form (4.1) to the solutions $I_S(1,2)$
and $I_S(2,2)$. As seen in Fig. 6, the maximum values of the fits
(4.1) are systematically well below the maximum values of the
experimental points for $I_S$. We conclude that the parametrization
(4.1) does not represent the experimental data very well. The likely
reason for this is the sudden drop of values of $I_S$ at 790 and
810 MeV which can be traced back to the dip in $|S|^2\Sigma$ at 790
MeV (see Fig. 5).

The results of the fits to the $S$-wave intensity $I_S$ at 5.98
GeV/c are presented in Table 3 and shown in Figure 7. Again, the
maximum values of the Breit-Wigner fits (4.1) are systematically
below the maximum experimental points of $I_S$. Because of large
errors on $I_S$ at this energy, the resonance parameters in Table 3
have also larger errors.
\vfill\eject
\noindent
{\bf 5. SPIN DEPENDENCE OF $\rho^0$ PRODUCTION}
\medskip
Resolution of the $\rho^0$ peak seen in the reaction cross-section
$d^2\sigma/dmdt$ into its individual spin components (8 moduli of
nucleon transversity amplitudes) is shown in Figures 3 and 4 at 17.2
and 5.98 GeV/c, respectively. We stress that this model independent
resolution is possible only in measurements on polarized targets.
We discussed the $S$-wave contributions in the preceeding Section.
In this Section we focus on the $P$-wave contributions. We observe
several important features in the spin dependence of $\rho^0$
production on the level of amplitudes.

\noindent
(a) The mass spectrum on the level of amplitudes depends on dimeson
helicity $\lambda$ and on the spin of the recoil nucleon. At low
momentum transfer (Fig. 3) the unnatural exchange amplitudes
$|\overline L |^2\Sigma$ and $|\overline U|^2\Sigma$ with recoil
nucleon spin up are larger than the amplitudes $|L|^2\Sigma$ and
$|U|^2\Sigma$ with recoil nucleon spin down. The opposite is true
for natural exchange amplitudes $|\overline N|^2\Sigma$ and
$|N|^2\Sigma$. These findings are similar to observations reported
in Fig.~10 of Ref.~4. At larger momentum transfers (Fig. 4) we also
find $|\overline L|^2\Sigma$ and $|N|^2\Sigma$ larger than
$|L|^2\Sigma$ and $|\overline N|^2\Sigma$, respectively, but
$|\overline U|^2\Sigma\approx |U|^2\Sigma$. These large differences
between the unnatural amplitudes with recoil nucleon spin up and
down are due to the contributions from large and nontrivial
$A_1$-exchange amplitudes $L_0$ and $U_0$. The same interpretation
was given in Ref.~4 and 5. We note that the $t$-dependence of
amplitudes in the $\rho^0$ mass region also shows evidence for large
$A_1$ exchange amplitudes (see Fig.~7 and 15 of Ref.~4, and Fig.~1
and 3 of Ref.~12).

\noindent
(b) We find unexpected suppression of $\rho^0$ production in
several amplitudes. At low momentum transfer (Fig. 3) the moduli
$|\overline N|^2\Sigma$ and $|U|^2\Sigma$ are small and flat and do
not show the expected $\rho^0$ peak. Also, while the amplitude
$|L|^2\Sigma$ is still large, it does not show a clear $\rho^0$ peak
but rather a broad structure. At the larger momentum transfers
(Fig. 4) we find $\rho^0$ suppression in the amplitudes $|\overline
U|^2\Sigma$ and $|U|^2\Sigma$ which are small relative to the other
$P$-wave amplitudes.$^{23}$

\noindent
(c) The mass spectra of the dominant amplitudes $|\overline
L|^2\Sigma$ and $|L|^2\Sigma$ show unexpected narrow structures
within the $\rho^0$ mass region which are not seen in the
spin-averaged cross-section $I_L$. The $P$-wave intensities $I_L$,
$I_N$ and $I_U$ are shown in Fig. 9 and 10 for 5.98 and 17.2 GeV/c,
respectively. We notice the narrow range of values in these
partial-wave intensities compared to the range of the moduli. This
indicates that the values of moduli with recoil nucleon spin up and
down are highly correlated for each $P$-wave amplitude.

First we look at the amplitudes at 5.98 GeV/c (Fig. 9). The mass
spectrum of the recoil nucleon spin up amplitude $|\overline
L|^2\Sigma$ shows a clear dip at 757 MeV and a peak at 807 MeV. The
opposite behaviour is seen in the nucleon spin down amplitude
$|L|^2\Sigma$ which peaks at 757 MeV and has a dip at 807 MeV. These
spin correlated structures within $\rho_0$ mass region do not
appear in the partial-wave intensity $I_L$ which shows a
structureless $\rho^0$ line shape.

Next we note a similar situation in the natural exchange amplitudes
but at a different mass. The spectrum of $|N|^2\Sigma$ shows a
maximum at 782 MeV which is associated with a pronounced dip in
$|\overline N|^2\Sigma$ at the same mass. The partial wave intensity
$I_N$ again shows a structureless line shape expected from a
$\rho^0$

One should note however that the apparently most significant
($2-3\sigma$) narrow structure at 807 MeV is due to a single
deviating date point in the $\rho^y_{1-1} - \rho^y_n$ (see Fig.~8 of
Ref.~10). This deviation is not observed in the $\rho^y_{ss} +
\rho^y_{00} + 2\rho^y_{11}$, also involving the difference between
$|L|^2$ and $|\overline L|^2$. This shows that the structures at
5.98 GeV/c can be treated as indications only.

Partial-wave intensity $I_L$ at 17.2 GeV/c at lower momentum
transfer (Fig. 10) shows a structureless line shape which peaks at
790 MeV. The expectation that the same structureless line shape is
reproduced on the level of amplitudes is not met again. Instead we
find spin correlated structures in the amplitude $|\overline
L|^2\Sigma$ and $|L|^2\Sigma$. The amplitude $|\overline L|^2\Sigma$
peaks at 790 MeV and has a lower value at 770 MeV. The opposite is
seen in the amplitude $|L|^2\Sigma$ which has a lower value at 790
MeV and peaks at 770 MeV.

The partial-wave intensities presented in Fig. 9 and 10 are
solution $I_A (1,1),\ A=L,U,N$. The other solutions are similar.

At this point one may ask how significant is the evidence for the
structures in the $\rho$ line shape on the level of amplitudes. We
note that the structures occur in $(m,t)$ bins with relatively high
statistics which provides a measure of confidence in the results.
Also, the $t$-evolution of mass dependence of moduli of normalized
amplitudes shows $t$-dependent structures within $\rho^0$ mass
region (see Fig.~1 of Ref.~13). Another possibility is to fit
Breit-Wigner form (4.1) to the spin-averaged partial wave intensity
$I_L$ and superimpose this Breit-Wigner fit (suitably normalized)
over the mass spectra in moduli $|\overline L|^2 \Sigma$ and
$|L|^2\Sigma$. The numerical results of the fits to $I_L$ are given
in Tables 4 and 5. In Fig.~11 we show the results for reaction
$\pi^+ n \to \pi^+\pi^- p$ at 5.98 GeV/c and $-t = 0.2 - 0.4$
(GeV/c).$^2$ The Breit-Wigner fit to $I_L$ peaks at 780 MeV.
Superimposed over the moduli $|\overline L |^2\Sigma$ and $|L
|^2\Sigma$, the Breit-Wigner fit shows that the narrow structures in
their mass spectra are statistically significant. At 757 MeV, the
structures are $1\sigma$ effect (dip in $|\overline L |^2\Sigma$ and
peak in $|L|^2\Sigma$). At 807 MeV, the structures are 2--3$\sigma$
effect (peak in $|\overline L |^2\Sigma$ and dip in $|L|^2\Sigma$).
In Fig.~10 we show the results for $\pi^- p \to \pi^- \pi^+ p$ at
17.2 GeV/c and $-t = 0.005 - 0.2$ (GeV/c).$^2$ While the
Breit-Wigner form fits the intensity $I_L$ very well we see that the
mass spectrum of amplitude $|\overline L |^2\Sigma$ is narrower
than the Breit-Wigner fit and the mass spectrum of amplitude
$|L|^2\Sigma$ is broader than the Breit-Wigner fit to $I_L$. At
present, the evidence for structures in the $\rho$-line shape of
amplitudes  at larger $t$ is still preliminary and insufficient for
theoretical analyses. New experiments on polarized targets with
significantly higher statistics than the 60,000 events (at 5.98
GeV/c) in the Saclay measurement$^{10}$ are required to confirm the
existence of such structures and to investigate the $t$-dependence
of $\rho$ line shape on the level of amplitudes.$^{13}$

We now briefly discuss the physical significance of the observed
spin effects in $\rho^0$ production. The presence of $A_1$-exchange
(feature (a)) will be discussed in the next Section. Here we will
focus on the features (b) and (c).

Following the discovery of resonances in hadron interactions, it
has always been believed that the production and decay of resonances
were separate events. For instance, the reaction $\pi^+n \to
\pi^+\pi^- p$ was thought of as a two step process in the $\rho^0$
mass region: $\rho^0$ production $\pi^+ n \to \rho^0 p$ followed by
$\rho^0$ decay $\rho^0 \to \pi^+\pi^-$. In this picture the
$S$-matrix elements factorize into production and decay matrix
elements
$$T(\pi^+ n \to \pi^+\pi^- p) = T (\pi^+ n \to \rho^0 p) \phi
(\rho^0) T (\rho^0 \to \pi^+ \pi^-)\eqno(5.1)$$
\noindent
where $\phi(\rho^0)$ is a $\rho^0$ propagator leading to the
Breit-Wigner description of dipion mass dependence of modulus of
each production amplitude $|T (\pi^+ n \to \pi^+ \pi^- p)|^2$ as
well as the production cross-section$^{24}$ $\Sigma = d^2
\sigma/dmdt$.

It is expected from (5.1) that the same $\rho^0$ resonance line
shape seen in the spin-averaged cross section $d^2\sigma/dmdt$ will
appear also in the modulus of every $P$-wave spin-dependent
production amplitude. The suppression of $\rho^0$ production
observed in several spin amplitudes and described above in part (b)
seems to invalidate the factorization hypothesis and the underlying
simple picture of resonance production. However, there is no
theoretical explanation for the suppression of $\rho^0$ production
in these spin amplitudes (Fig. 3 and 4). Evidently, the spin
amplitudes showing $\rho_0$ suppression cannot be fitted with
Breit-Wigner form.

The factorization hypothesis (5.1) also implies that the line shape
of the mass spectrum of the decaying resonance does not depend on
the nucleon spin and on the momentum transfer $t$. However, the line
shapes of the dominant amplitudes $|\overline L|^2\Sigma$ and
$|L|^2\Sigma$ at larger $t$ show unexpected structures within the
$\rho^0$ mass region which correlate the mass spectra corresponding
to opposite nucleon spins. Comparison of the structures in Fig. 9
and 10 indicates a change as we go from low to larger momentum
transfer. This suggests that the narrow dips and peaks observed in
these moduli within the $\rho^0$ mass region evolve with $t$ (see
also Ref.~13). This $t$-dependence of the resonance line shape in
the amplitudes is also at variance with the factorization
hypothesis.

The suppression of $\rho^0$ production and the narrow structures
observed in dominant amplitudes in $\rho^0$ mass region represent a
new information on pion production and raise questions about the
nature of hadron resonances. A resonance which is a pole in the
amplitude would lead to simple resonance peaks in all moduli without
structures within their widths. We also note that the usual
models$^{25}$ of meson resonances as $q\overline q$ bound states do
not predict their hadronic widths and line shape.
\bigskip
\noindent
{\bf 6. TESTS OF ASSUMPTIONS USED IN DETERMINATIONS OF $\pi\pi$
PHASE SHIFTS.}
\medskip
We have presented a model independent and solution independent
evidence for the scalar state $I=0\ 0^{++} (750)$ in measurements of
$\pi N_\uparrow \to \pi^+\pi^- N$ on polarized target. The question
arises how to understand the absence of such a state in the
accepted solution for $S$-wave phase shift $\delta_0^0$ in $\pi\pi$
scattering.$^{\rm 2, 26-29}$

Of course, there are no actual measurements of pion-pion scattering
and there is no partial-wave analysis of $\pi\pi \to \pi\pi$ in the
usual sense. The $\pi\pi$ phase shifts are determined indirectly
from measurements of $\pi^- p \to \pi^-\pi^+ n$ on unpolarized
target using extrapolations into unphysical region of momentum
transfer $t$ and several necessary enabling assumptions. Some of
these crucial assumptions lead to predictions for polarized
spin-density-matrix (SDM) elements and are thus directly testable in
measurements on polarized targets. As we shall see in detail below,
these assumptions are invalidated in a major way by the data on
polarized targets. We must use the results of measurements on
polarized targets to judge the validity of $\pi\pi$ phase shift
determinations, and not vice versa. We are thus led to the
conclusion, that the indirect and model dependent determinations of
$\pi\pi$ phase shifts cannot be correct. This explains the absence
of $I=0\ 0^{++} (750)$ resonance in $\delta_0^0$ phase shift from
these analyses.

To unveil the assumptions used in determinations of $\pi\pi$ phase
shifts we will trace step by step the procedure used by Estabrooks
and Martin.$^{27, 28}$ Their approach has the advantage of being the
most transparent and of using the least number of assumptions.
Moreover, their assumptions are common to all other determinations
of $\pi\pi$ phase shifts from unpolarized $\pi N \to \pi^+\pi^- N$
data.$^{2, 26, 29}$ We shall restrict our discussion to dipion
masses below 1000 MeV where $S$- and $P$-wave dominate.

The starting point are dimeson helicity $\lambda=0$ pion exchange
amplitudes $S_1$ and $L_1$ in the $t$-channel. It is assumed that
the $t$- and $m$-dependence in these amplitudes factorizes$^{27,
28}$
$$\eqalignno{S_1 (m,t)& = {{\sqrt{-t}}\over{t-\mu^2}} F_0 (t)
{m\over{\sqrt q}} f_0 (m)&(6.1)\cr
L_1 (m,t) &= {{\sqrt{-t}}\over{t-\mu^2}} F_1(t) {m\over{\sqrt q}}
\sqrt 3 f_1 (m)\cr}$$
\noindent
where $t$ is the momentum transfer at the nucleon vertex, $m$ and
$q$ are dipion mass and $\pi^-$ momentum in the $\pi^-\pi^+$ c.m.
frame. The form factors $F_J(t)$ describe the $t$-dependence and the
functions $f_J(m), J=0,1$ describe the mass dependence. Further,
the functions $f_J(m)$ are assumed to be partial wave amplitudes in
$\pi^-\pi^+ \to \pi^-\pi^+$ reaction at c.m. energy $m$
$$\eqalignno{f_0&={2\over 3} f_0^{I=0} + {1\over 3} f_0^{I=2}&(6.2)\cr
f_1&= f_1^{I=1}\cr}$$
\noindent
The partial wave amplitudes $f^I_J$ with definite isospin $I$ are
defined so that in the $\pi\pi$ elastic region
$$f^I_J = \sin\delta^I_J e^{i\delta^I_J}\eqno(6.3)$$
\noindent
The $\pi\pi$ phase shifts $\delta^I_J$ can be obtained by
extrapolating the production amplitudes $S_1$ and $L_1$ from the
physical region $t < 0$ to $t = \mu^2$. One cannot determine both
$I=0$ and $I=2$ $S$-wave phase shifts and so values for $f^2_0$
obtained in analyses of $\pi^+ p \to \pi^+ \pi^+ n$ data were
used.$^{27, 28}$

There is no theoretical proof of factorization (6.1) and
identification (6.2) of functions $f_J$ with $\pi\pi$ partial wave
amplitudes. Strictly speaking, the relations (6.1)--(6.3) are
definitions of $\pi\pi$ phase shifts. It is by no means obvious that
these phase shifts would coincide with $\pi\pi$ phase shifts
determined directly from real pion-pion scattering. Only such
comparison would test the assumption (6.2) which is not testable in
measurements of $\pi N \to \pi^+\pi^- N$.

Essential element in the determination of $\pi\pi$ phase shifts is
the knowledge of the production amplitudes $S_1$ and $L_1$. But an
inspection of equations (2.8a) reveals that the amplitudes $S_1$ and
$L_1$ cannot be calculated from the data on unpolarized target
without additional assumptions. There are simply more amplitudes
than data. To proceed further all determinations of $\pi\pi$ phase
shifts must assume that all $A_1$-exchange amplitudes vanish, i.e.
$$S_0 = L_0 = U_0 \equiv 0\eqno(6.4)$$
\noindent
With the assumptions (6.4) and notation (2.9) the equations (2.8a)
now read
$$\eqalignno{c_1&=\rho_{ss} + \rho_{00} + 2\rho_{11} = |S_1|^2 +
|L_1|^2 + |U_1|^2 + \sigma_N \equiv 1&(6.5)\cr
c_2&= \rho_{00} - \rho_{11} = |L_1|^2 - {1\over 2} |U_1|^2 -
{1\over 2} \sigma_N\cr
c_3&= \rho_{1-1} = {1\over 2} \sigma_N -{1\over 2} |U_1|^2\cr
c_4&= \sqrt 2 Re \rho_{10} = Re (U_1 L_1^*) = |U_1| |L_1| \cos
(\chi_{\lower.5ex\hbox{$\scriptstyle {LU}$}})\cr
c_5&= \sqrt 2 Re \rho_{1s} = Re (U_1 S_1^*) = |U_1| |S_1| \cos
(\chi_{\lower.5ex\hbox{$\scriptstyle {SU}$}})\cr
c_6&= Re \rho_{0s} = Re (L_1 S_1^*) = |L_1| |S_1| \cos
(\chi_{\lower.5ex\hbox{$\scriptstyle {SL}$}})\cr}$$
\noindent
where the relative phases satisfy identity
$$\chi_{\lower.5ex\hbox{$\scriptstyle {LU}$}} -
\chi_{\lower.5ex\hbox{$\scriptstyle {SU}$}} +
\chi_{\lower.5ex\hbox{$\scriptstyle {SL}$}} =
(\phi_{\lower.5ex\hbox{$\scriptstyle {L_1}$}} -
\phi_{\lower.5ex\hbox{$\scriptstyle {U_1}$}}) -
(\phi_{\lower.5ex\hbox{$\scriptstyle {S_1}$}} -
\phi_{\lower.5ex\hbox{$\scriptstyle {U_1}$}}) +
(\phi_{\lower.5ex\hbox{$\scriptstyle {S_1}$}} -
\phi_{\lower.5ex\hbox{$\scriptstyle {L_1}$}})=0\eqno(6.6)$$
\noindent
The equations (6.5) are formally similar to the set (2.13) and can
be similarly solved. We obtain
$$\eqalignno{|S_1|^2& = 1 + 2 c_2 - 3|L_1|^2&(6.7)\cr
|U_1|^2 &= |L_1|^2 - (c_2 + c_3)\cr
\sigma_N&= |L_1|^2 - (c_2 - c_3)\cr
\cos\chi_{\lower.5ex\hbox{$\scriptstyle {LU}$}} &=
{c_4\over{|L_1||U_1|}}\cr
\cos\chi_{\lower.5ex\hbox{$\scriptstyle {SU}$}}& =
{c_5\over{|S_1||U_1|}}\ ,\  \cos \chi_{\lower.5ex\hbox{$\scriptstyle
{SL}$}}= {c_6\over{|S_1||L_1|}}\cr}$$
\noindent
For $|L_1|^2 \equiv x$ we have a cubic equation
$$a x^3 + bx^2 + cx + d = 0\eqno(6.8)$$
\noindent
where
$$\eqalign{a&=3\cr
b=&(1+ 5 c_2 + 3 c_3)\cr
c=& (1+ 2 c_2) (c_2 + c_3 ) - 3 c^2_4 + c_5^2 + c_6^2\cr
d=& (1+2c_2) c_4^2 + (c_2 + c_3) c_6^2 - 2 c_4 c_5 c_6\cr}$$

Using the data on unpolarized target at 17.2 GeV/c [Ref. 2],
Estabrooks and Martin found$^{27, 28}$ in all $(m,t)$ bins 3 real
solutions. One of the solutions is always negative and it is
rejected. The other two solutions for $|L_1|^2$ yield two solutions
for $|S_1|^2$. Thus there are two solutions for $S$- and $P$-wave
phase shifts below 1000 MeV. Solution 1 for $\delta_0^0$ is
nonresonating while the solution 2 resonates at the mass around 770
MeV with a width about 150 MeV. The resonating solution was rejected
because it disagreed with a $\pi^0\pi^0$ mass spectrum from a low
statistics experiment$^{30}$ on $\pi^- p \to \pi^0\pi^0 n$ at 8
GeV/c.

The absence of $A_1$ exchange amplitudes is absolutely crucial for
the determination of $\pi\pi$ phase shifts from unpolarized target
data on $\pi^- p \to \pi^- \pi^+ n$. Fortunately, it is an
assumption that can be tested directly in experiments on polarized
targets. Using the definitions (2.4) of nucleon transversity
amplitudes, the assumptions (6.4) imply that the moduli of unnatural
exchange amplitudes with recoil nucleon spin up and down must be
equal, i.e.
$$|\overline S|^2 = |S|^2,\quad |\overline L|^2= |L|^2,\quad
|\overline U|^2 = |U|^2\eqno(6.9)$$
\noindent
These predictions are in sharp disagreement with the results of
model independent amplitude analysis of the data on polarized target
at 17.2 GeV/c (Fig. 3) which finds
$$|\overline S|^2 > |S|^2,\quad |\overline L|^2 > |L|^2,\quad
|\overline U|^2 > |U|^2\eqno(6.10)$$
\noindent
Particularly important is the large difference between $|\overline
L|^2$ and $|L|^2$ (see also Fig. 10) which can be accounted for only
by a large and nontrivial $A_1$-exchange contribution from the
amplitude $L_0$. The assumptions (6.4) are thus badly violated. This
implies that the determination of amplitudes $|L_1|^2$ and
$|S_1|^2$ through equations (6.7) and (6.8) cannot be correct. This
in turn means through (6.1) that the $\pi\pi$ phase shifts cannot be
correctly determined from the data on unpolarized target.

Using the assumptions (6.4) in equations (2.8b) leads to the
following predictions for polarized SDM elements:
$$\rho_{ss}^y + \rho_{00}^y + 2 \rho_{11}^y = -2 (\rho_{00}^y -
\rho_{11}^y) = + 2 \rho_{1-1}^y\eqno(6.11)$$
$$Re \rho_{10}^y = Re \rho_{1s}^y = Re \rho_{0s}^y \equiv 0\eqno(6.12)$$
\noindent
The data for polarized SDM elements clearly rule out these
predictions as is shown in Fig.~12 and 13 for $\pi^- p \to \pi^-
\pi^+ n$ reaction at 17.2 GeV/c. We find that $\rho_{ss}^y +
\rho_{00}^y + 2 \rho_{11}^y$ and $-2 (\rho_{00}^y - \rho_{11}^y)$
have large magnitudes but opposite signs while $2\rho_{1-1}^y$ has a
small magnitude. The interference terms $Re \rho_{10}^y,
Re\rho_{1s}^y$ and $Re \rho_{0s}^y$ are all dissimilar and have
large non zero values. On the basis of this evidence we again must
conclude that the determinations of $\pi\pi$ phase shifts from
unpolarized data on $\pi N \to \pi^+ \pi^- N$ are questionable.

If the pion exchange amplitudes $S_1$ and $L_1$ cannot be reliably
determined from the data on unpolarized targets, the question arises
whether these amplitudes can be determined from the data on
polarized targets. Unfortunately the data on polarized target do not
allow the separation of $\pi$- and $A_1$-exchange amplitudes. In
fact it has been shown recently$^{31}$ that the pion exchange
amplitudes $S_1$ and $L_1$ can be expressed in terms of data on
polarized target plus the $A_1$-exchange amplitudes $S_0$ and $L_0$.
These relations show explicitly that the determination of $\pi\pi$
phase shifts from data on polarized target depends on the model used
for $A_1$-exchange amplitudes $S_0$ and $L_0$. We conclude that in
the absence of a reliable model for $A_1$-exchange or real pion-pion
scattering data the $\pi\pi$ phase shifts remain undetermined.

It is of interest to compare our results for $S$-wave intensity
with previous $\pi\pi$ phase shift analyses to assess the effect of
$A_1$ exchange and the contribution of isospin $I=2$ $S$-wave
amplitudes.

In Fig.~14 and 15 we compare $S$-wave cross-sections normalized to
one at maximum value. The data points correspond to $I_S(1,1)$ and
$I_S(2,2)$ at 17.2 GeV/c in the physical region of $t$. The curves
in Fig.~14 (taken from Ref.~15) are predictions of various
determinations of $\pi\pi$ phase shifts. The curves $A$ and $D$ are
the two solutions of Estabrooks and Martin (Ref.~27). The curves $B$
and $C$ are predictions from Ref.~32 and 33, respectively. The
predictions $A$, $B$ and $C$ show broad structures in $I_S$ around
750 MeV where the data on polarized target require a narrower
structure. The broad structure of predictions $A$, $B $ and $C$ does
not agree well with the current algebra prediction above threshold
(curve $E$, Ref.~34).

The solid curves in Fig.~15 are prediction from a recent fit to
$\delta^0_0$ phase shift data by Zou and Bugg.$^{35}$ Results from
another recent fit by Au, Morgan and Pennington$^{36}$ are very
similar. The predictions are broad and peak at 860 MeV while our
data are much narrower and peak at 750--770 MeV. The predictions
from phase shift analyses$^{35,36}$ were calculated using $I_S =
|f_0|^2$ where $f_0 = {2\over 3} f_0^{I=0}$ with $f_0^{I=0}$ given
by (6.3).

All curves in Fig.~14 and the solid curves in Fig.~15 correspond to
$I=0$ $S$=wave intensity in $\pi^+\pi^- \to \pi^+ \pi^-$ reaction.
The data points in Fig.~14 and 15 correspond to $S$-wave intensity
in $\pi^- p \to \pi^- \pi^+ n$ at 17.2 GeV/c in the physical region
of $t$ and include contribution from isospin $I=2$ $S$-wave
amplitudes. The question arises to what extend the $I=2$ $S$-wave
amplitudes affect the resonance shape of $\delta (750)$ state seen
in the $S$-wave intensity data $I_S(1,1)$. We cannot answer this
question using the data on polarized target since these data do not
allow isolation of the two $I=2$ $S$-wave amplitudes. However, we
can use (6.2) and the data for $I=2$ phase shift $\delta_0^2$ to
calculate the $S$-wave intensity $I_S = |f_0|^2$ in $\pi^+\pi^-$
scattering with the $I=2$ contribution.

The experimental information about $I=2$ $S$-wave phase shifts
comes from the analysis of $\pi^+ p \to \pi^+ \pi^+ n$ reaction. We
used the results from the 1977 analysis of Hoogland et al. in
Ref.~37. This analysis presents two very similar results for
$\delta_0^2$ based on two different methods (method A and method B).
The dashed line in Fig.~15 shows the $S$-wave intensity of Zou and
Bugg with the $I=2$ correction using $\delta_0^2$ from method A in
Ref.~37. The method B yields essentially identical curve. We see
that the $I=2$ correction maintains the overall shape of $S$-wave
intensity in $\pi^+\pi^- \to \pi^+ \pi^-$ and does not secure
agreement with the data from $\pi^- p \to \pi^- \pi^+ n$ analysis.
We can conclude that the principal cause of the difference between
the $\pi\pi$ phase shift analyses and our amplitude analysis of
$\pi^- p \to \pi^- \pi^+ n$ data on polarized target is the
existence of large $A_1$ exchange amplitudes in this reaction.
\vfill\eject
\noindent
{\bf 7. SUMMARY}
\medskip
The measurements of reactions $\pi^- p_\uparrow \to \pi^- \pi^+ n$
at 17.2 GeV/c and $\pi^+ n_\uparrow \to \pi^+ \pi^- p$ at 5.98 and
11.85 GeV/c on polarized targets enable a model-independent
amplitude analysis for dipion masses below 1000 MeV where $S$-wave
and $P$-wave contributions dominate the pion production process. The
amplitude analysis yields two similar solutions for 8 moduli and 6
cosines of relative phases. In most $(m,t)$ bins the analytical
solution produces unphysical values of some cosines or moduli,
violating the conditions (2.17). These conditions are equivalent to
inequalities constraints on SDM elements.$^{12, 38}$ To avoid
unphysical solutions the constraints should be imposed during the
optimization of maximum likelihood function in data tapes analysis
in future experiments. This will require the use of methods for
constrained optimization.$^{16, 39, 40, 41}$

The previous amplitude analyses$^{14}$ found evidence for a new
scalar state $I=0\ 0^{++}(750)$ which can be interpreted as the
lowest mass scalar gluonium $0^{++}(gg).^{14}$ Because of the
significance of this finding it is important to see what happens to
the scalar mass spectra when the contamination by unphysical
solutions is removed. To filter out the unphysical solutions we used
a Monte Carlo approach. The input SDM elements were randomly varied
within errors 30~000 times in each $(m,t)$ bin and amplitude
analysis was performed for each selection. Only when the 8 moduli
and 6 cosines attain physical values in both solutions are their
values retained and partial-wave intensities and polarizations are
calculated. All such physical solutions are collected and these
collections then define the distribution, range and average values
for each modulus, cosine of relative phase, partial wave intensity
and polarization.

After filtering out the unwanted unphysical solutions, a clear
signal for $I=0\ 0^{++}(750)$ resonance emerges in all solutions for
partial wave intensity $I_S$ at larger momentum transfers at 5.98
and 11.85 GeV/c, and in two solutions at lower momentum transfers at
17.2 GeV/c. The ambiguity at 17.2 GeV/c is caused by the non
resonating amplitude $|S|^2\Sigma$ which masks the resonating
behaviour of $|\overline S|^2\Sigma$.  The mass spectra in $I_S$
peak at or near 750 MeV. The width at half-height is about 230 MeV
and 270 MeV at 17.2 and 5.98 GeV/c, respectively. At 11.85 GeV/c,
the solutions $I_S(2,1)$ and $I_S(2,2)$ are consistent with a narrow
width of 150 MeV. At 17.2 GeV/c, the amplitude $|\overline
S|^2\Sigma$ clearly resonates in both solutions at 750--770 MeV with
a width at half-height estimated to be 175 MeV. The production of
$I=0\ 0^{++}(750)$ appears suppressed in the amplitude
$|S|^2\Sigma$. There is a clear phase degeneracy between amplitudes
$\overline S$ and $\overline L$, and $S$ and $L$, also indicating
resonant structure of the $S$-wave near $\rho^0$ mass.

The results for $\rho^0$ production generally confirm the findings
of previous analyses.$^{4, 12, 13, 14}$ We find significant
suppression of $\rho^0$ production in amplitudes $|\overline
N|^2\Sigma$ and $|U|^2\Sigma$ at all energies. There is an
additional $\rho^0$ suppression in $|L|^2\Sigma$ at 17.2 GeV/c and
in $|\overline U|^2\Sigma$ at 5.98 GeV/c. Moreover, the line shapes
of $|\overline L|^2 \Sigma$ and $|L|^2\Sigma$ show unexpected
structures within $\rho^0$ mass region which correlate the mass
spectra corresponding to opposite nucleon spins. These narrow
structures are not observed in the spin averaged partial wave
intensities $L_L$ which show the expected structureless $\rho^0$
peak. There is no theoretical explanation for these important
features of $\rho^0$ production which contradict the factorization
hypothesis.

All determinations of $\pi\pi$ phase shifts from unpolarized data
on $\pi^- p \to \pi^- \pi^+ n$ rely on the absence of $A_1$-exchange
amplitudes. This essential assumption leads to predictions for
polarized SDM elements which are clearly ruled out by the data on
polarized target. In our Monte Carlo amplitude analysis we find a
clear evidence for large and nontrivial $A_1$-exchange contributions
from $S$-wave and $P$-wave amplitudes, in particular at 17.2 GeV/c.
On the basis of this evidence we come to the conclusion that the
usual determinations of $\pi\pi$ phase shifts cannot be correct.
This also explains why the accepted solution for $\delta_0^0$ phase
shift shows no evidence for the $\sigma(750)$ resonance.

Production of scalar state $I=0\ 0^{++}(750)$, the suppression of
$\rho^0$ production in certain production amplitudes and the
possible narrow structures in moduli in $\rho^0$ mass region are
unexpected and important findings of measurements of $\pi N \to
\pi^+ \pi^- N$ reactions on polarized targets. These new phenomena
should be further investigated experimentally in a new generation of
dedicated high statistics experiments with polarized targets.$^{42,
43}$
\bigskip
\noindent
{\bf Acknowledgments}
\medskip
I wish to thank J.G.H. de Groot for numerical results of Ref. 4 and
A. de Lesquen for useful discussions and help with the program
FUMILI. This work was supported by Fonds pour la Formation de
Chercheurs et l'Aide \`a la Recherche (FCAR), Minist\`ere de
l'Education du Qu\'ebec, Canada.
\vfill\eject
\noindent
{\bf References}
\medskip
\item{[1]} G. Lutz and K. Rybicki, Max Planck Institute, Munich,
Internal Report No. MPI-PAE/Exp. EI.75, 1978 (unpublished).
\item{[2]} G. Grayer et al., Nucl. Phys. \underbar{B75}, 189 (1974).
\item{[3]} J.G.H. de Groot, PhD Thesis, University of Amsterdam,
1978 (unpublished).
\item{[4]} H. Becker et al., Nucl. Phys. \underbar{B150}, 301 (1979).
\item{[5]} H. Becker et al., Nucl. Phys. \underbar{B151}, 46 (1979).
\item{[6]} V.~Chabaud et al., Nucl. Phys. \underbar{B223}, 1 (1983).
\item{[7]} K.~Rybicki, I.~Sakrejda, Z. Phys. \underbar{C28}, 65 (1985).
\item{[8]} See Fig. 2 of Ref. 5 and Fig.~1 of Ref.~6.
\item{[9]} M. Babou et al., Nucl. Instrum. Methods \underbar{160},
1 (1979).
\item{[10]} A. de Lesquen et al., Phys. Rev. \underbar{D32}, 21 (1985).
\item{[11]} A. de Lesquen et al., Phys. Rev. \underbar{D39}, 21 (1989).
\item{[12]} M. Svec, A. de Lesquen and L. van Rossum, Phys. Rev.
\underbar{D45}, 55 (1992).
\item{[13]} M. Svec, A. de Lesquen and L. van Rossum, Phys. Rev.
\underbar{D42}, 934 (1990).
\item{[14]} M. Svec, A. de Lesquen and L. van Rossum, Phys. Rev.
\underbar{D46}, 949 (1992). An error in $d^2\sigma/dmdt$ at 17.2
GeV/c occurred in this paper. The results for moduli and intensities
at 17.2 GeV/c are inaccurate and the conclusion that all four
solutions for $S$-wave intensity at 17.2 GeV/c resonate at 750 MeV
is not valid.
\item{[15]} N.M. Cason et al., Phys. Rev. \underbar{D28}, 1586 (1983).
\item{[16]} R.K. Brayton, R. Spence, Sensitivity and Optimization
(Elsevier Scientific, Amsterdam, 1980) p. 29.
\item{[17]} F. James, in Techniques and Concepts of High Energy
Physics II (Plenum Press, New York, 1983), Ed. Thomas Ferbel, p.
196.
\item{[18]} H.~Palka, Institute of Nuclear Physics, Cracow,
Internal Report No.~1230/PH, Appendix D, 1983 (unpublished).
\item{[19]} Subroutine SURAND, in IBM Engineering and Scientific
Subroutine Library, Guide and Reference, Release 4, Fifth Edition
March 1990, p. 852.
\item{[20]} In our computer program the order of observables is the
same as in (2.8a) and (2.8b) except the order of $\rho_{1s},
\rho_{0s}$ and $\rho_{1s}^y, \rho_{0s}^y$ is reversed to $\rho_{0s},
\rho_{1s}$ and $\rho_{0s}^y, \rho_{1s}^y$.
\item{[21]} I. Silin, FUMILI Long Write-up D510, CERN Computer
Centre Program Library, revised 1985.
\item{[22]} J.M. Blatt, V.F. Weiskopf, Theoretical Nuclear Physics,
Wiley \& Sons, New York, 1952.
\item{[23]} Notice in Figure 9 that the intensity $I_U$ shows a
bump at $\rho^0$ mass which is smaller than the $\rho^0$ peaks in
the other intensities $I_L$ and $I_N$. This structure in $I_U$ is
consistent with the results shown in Fig.~10 of Ref.~7.
\item{[24]} H. Pilkuhn, The Interactions of Hadrons, North-Holland
1967, p. 33.
\item{[25]} D. Flamm and F. Sch\"obert, Introduction to quark model
of elementary particles, Gordon and Breach Science Publishers,
1982.
\item{[26]} W. Ochs, PhD Thesis, Ludwig-Maximilians-Universit\"at,
M\"unchen, 1973.
\item{[27]}  P. Estabrooks, A.D. Martin, $\pi\pi$ Scattering-1973,
Proceedings of the Int. Conf. on $\pi\pi$ Scattering, Tallahassee,
1973, edited by D.K. Williams and V. Hagopian, AIP Conf. Proc. No.
13 (AIP, New York, 1973), p. 37.
\item{[28]} P. Estabrooks and A.D. Martin, Nucl. Phys.
\underbar{B79}, 301 (1974).
\item{[29]} B. Hyams et al., Nucl. Phys. \underbar{B64}, 134 (1973).
\item{[30]} W.D. Apel et al., Phys. Lett. \underbar{41B}, 542 (1972).
\item{[31]} M. Svec, Phys. Rev. \underbar{D47}, 2132 (1993).
\item{[32]} S.D.~Protopopescu et al., Phys. Rev. \underbar{D7},
1279 (1973).
\item{[33]} P.~Estabrooks, A.D.~Martin, Nucl. Phys. \underbar{B95},
322 (1975).
\item{[34]} S.~Weinberg, Phys. Rev. Lett. \underbar{17}, 616 (1966).
\item{[35]} B.S.~Zou and D.V.~Bugg, Phys.~Rev. \underbar{D48},
R3948 (1993).
\item{[36]} K.L.~Au, D.~Morgan, M.R.~Pennington, Phys.~Rev.
\underbar{D35}, 1633 (1987).
\item{[37]} W.~Hoogland et al., Nucl.~Phys. \underbar{B126}, 109 (1977).
\item{[38]} There are several typing errors in eqs. (3.19) of ref.
12. The correct inequalities read
$$\displaylines{\qquad {1\over 4} (a_1 + a_2) \le 1\hfill\cr
\qquad {1\over 2} |a_3| \le {4\over 3} - [{1\over 3} (a_1 + a_2)
-{1\over 2} a_2]\hfill\cr
\qquad |a_4| \le \sqrt {1-{1\over 2} (a_2 + a_3)}\hfill\cr}$$
\item{} Also, in eqs. (2.16a) of ref.~12 $\rho_{00} + \rho_{11}$
should read $\rho_{00} - \rho_{11}$.
\item{[39]} Ph. E. Gill, W. Murray, and M.H. Wright, Practical
Optimization (Academic, New York, 1981).
\item{[40]} B.A. Murtagh and M.A. Saunders, MINOS 5.1 User's Guide,
Systems Optimization Laboratory Report No. SOL 83--20R, Stanford
University, 1987.
\item{[41]} W.T. Eadie et al., Statistical Methods in Experimental
Physics (North-Holland, Amsterdam, 1971), p. 159.
\item{[42]} J.R. Comfort, in Proceedings of a Workshop on Science
at the KAON Factory, TRIUMF, 1990, Editor D.R. Gill, vol. 2.
\item{[43]} M. Svec, in Proceedings of a Workshop on Future
Directions in Particle and Nuclear Physics at Multi-GeV Hadron Beam
Facilities, Brookhaven National Laboratory, 1993, Editor D.
Geesaman, p.~401.
\vfill\eject
\tabskip=1em plus2em minus.5em
\halign to \hsize{\hfil #\hfil&\hfil #\hfil&#\hfil&\hfil
#\hfil&\hfil #\hfil&\hfil #\hfil&\hfil #\hfil&\hfil #\hfil&\hfil
#\hfil\cr
&{\bf INPUT}&&&{\bf PASS}&&&{\bf FAIL}\cr
\noalign{\smallskip}
{\bf MASS}&{\bf MEAN}&{\bf ERROR}&{\bf MIN}&{\bf MAX}&{\bf
AVRG}&{\bf MIN}&{\bf MAX}&{\bf AVRG}\cr
\noalign{\medskip}
0.610&-0.702&0.096&-0.796&-0.606&-0.653&-0.798&-0.606&-0.703\cr
0.630&-0.620&0.085&-0.704&-0.535&-0.585&-0.705&-0.535&-0.624\cr
0.650&-0.762&0.078&-0.823&-0.808&-0.816&-0.840&-0.684&-0.761\cr
0.670&-0.596&0.067&-0.663&-0.529&-0.600&-0.663&-0.528&-0.594\cr
0.690&-0.535&0.057&-0.584&-0.479&-0.509&-0.592&-0.479&-0.534\cr
0.710&-0.592&0.050&-0.642&-0.542&-0.585&-0.642&-0.542&-0.591\cr
0.730&-0.610&0.046&-0.656&-0.564&-0.595&-0.656&-0.564&-0.610\cr
0.750&-0.588&0.043&-0.631&-0.547&-0.591&-0.631&-0.546&-0.587\cr
0.770&-0.489&0.043&-0.532&-0.447&-0.486&-0.532&-0.447&-0.488\cr
0.790&-0.557&0.043&-0.598&-0.514&-0.543&-0.599&-0.514&-0.555\cr
0.810&-0.411&0.046&-0.410&-0.366&-0.380&-0.457&-0.365&-0.410\cr
0.830&-0.436&0.053&-0.489&-0.383&-0.420&-0.489&-0.383&-0.436\cr
0.850&-0.461&0.057&-0.488&-0.404&-0.426&-0.518&-0.404&-0.460\cr
0.870&-0.340&0.060&-0.399&-0.280&-0.315&-0.401&-0.280&-0.342\cr
0.890&-0.372&0.064&--&--&--&-0.436&-0.308&-0.371\cr}
\bigskip
\bigskip
\noindent
{\bf Table 1.} The mass dependence of $\rho_{ss}^y + \rho_{00}^y +
2\rho_{11}^y$ in $\pi^- p \to \pi^- \pi^+ n$ at 17.2 GeV/c for
$-t=$0.005--0.2 (GeV/c).$^2$ The input mean values and errors are
compared with the ranges and average values of $\rho_{ss}^y +
\rho_{00}^y + 2\rho_{11}^y$ for pass and fail Monte Carlo selections
for $N_{tot} =$30~000.
\vfill\eject
\tabskip=1em plus2em minus.5em
\halign to \hsize{\hfil #\hfil&\hfil #\hfil&\hfil #\hfil&\hfil
#\hfil&\hfil #\hfil\cr
\noalign{\hrule\bigskip}
$I_S$&$m_\sigma$&$\Gamma_\sigma$&$N_S$&$\chi^2/d.o.f.$\cr
Solution&(MeV)&(MeV)\cr
\noalign{\bigskip\hrule\bigskip}
(1,1)&764$\pm$ 6&273 $\pm$ 24&1.97 $\pm$ 0.07&0.360\cr
\noalign{\bigskip\hrule\bigskip}
(2,1)&761$\pm$ 12&290 $\pm$ 54&2.16 $\pm$ 0.16&0.129\cr
\noalign{\bigskip\hrule}\cr}
\bigskip
\bigskip
\noindent
{\bf Table 2.} Results of the fits to two resonating solutions of
$S$-wave intensity $I_S$ in $\pi^- p \to \pi^- \pi^+ n$ at 17.2
GeV/c using the parametrization (4.1).
\vfill\eject
\tabskip=1em plus2em minus.5em
\halign to \hsize{\hfil #\hfil&\hfil #\hfil&\hfil #\hfil&\hfil
#\hfil&\hfil #\hfil\cr
\noalign{\hrule\bigskip}
$I_S$&$m_\sigma$&$\Gamma_\sigma$&$N_S$&$\chi^2/d.o.f.$\cr
Solution&(MeV)&(MeV)\cr
\noalign{\bigskip\hrule\bigskip}
(1,1)&732$\pm$ 17&244 $\pm$ 36&0.60 $\pm$ 0.12&0.913\cr
\noalign{\bigskip\hrule\bigskip}
(1,2)&700$\pm$ 33&300 $\pm$ 66&1.21 $\pm$ 0.22&0.126\cr
\noalign{\bigskip\hrule\bigskip}
(2,1)&740$\pm$ 32&296 $\pm$ 116&1.02 $\pm$ 0.29&0.204\cr
\noalign{\bigskip\hrule\bigskip}
(2,2)&711$\pm$ 20&300 $\pm$ 60&1.70 $\pm$ 0.21&0.238\cr
\noalign{\bigskip\hrule}\cr}
\bigskip
\bigskip
\noindent
{\bf Table 3.} Results of the fits to the four solutions of
$S$-wave intensity $I_S$ in $\pi^+ n \to \pi^+ \pi^- p$ at 5.98
GeV/c using the parametrization (4.1).
\vfill\eject
\tabskip=1em plus2em minus.5em
\halign to \hsize{\hfil #\hfil&\hfil #\hfil&\hfil #\hfil&\hfil
#\hfil&\hfil #\hfil&\hfil #\hfil\cr
\noalign{\hrule\bigskip}
$I_L$&$m_\rho$&$\Gamma_\rho$&$r$&$N_L$&$\chi^2/d.o.f.$\cr
Solution&(MeV)&(MeV)&(GeV$^{-1}$)\cr
\noalign{\bigskip\hrule\bigskip}
(1,1)&777$\pm$ 1&160 $\pm$ 2&4.9 $\pm$ 0.5&4.52 $\pm$ 0.03&2.918\cr
\noalign{\bigskip\hrule\bigskip}
(1,2)&777$\pm$ 1&157 $\pm$ 3&4.5 $\pm$ 0.7&4.50 $\pm$ 0.05&1.106\cr
\noalign{\bigskip\hrule\bigskip}
(2,1)&777$\pm$ 1&160 $\pm$ 3&4.8 $\pm$ 0.7&4.50 $\pm$ 0.04&1.551\cr
\noalign{\bigskip\hrule\bigskip}
(2,2)&777$\pm$ 1&157 $\pm$ 3&4.3 $\pm$ 0.6&4.48 $\pm$ 0.04&1.596\cr
\noalign{\bigskip\hrule}\cr}
\bigskip
\bigskip
\noindent
{\bf Table 4.} Results of the fits to the four solutions of
$P$-wave intensity $I_L$ in $\pi^- p \to \pi^- \pi^+ n$ at 17.2
GeV/c using the parametrization (4.1).
\vfill\eject
\tabskip=1em plus2em minus.5em
\halign to \hsize{\hfil #\hfil&\hfil #\hfil&\hfil #\hfil&\hfil
#\hfil&\hfil #\hfil&\hfil #\hfil\cr
\noalign{\hrule\bigskip}
$I_L$&$m_\rho$&$\Gamma_\rho$&$r$&$N_L$&$\chi^2/d.o.f.$\cr
Solution&(MeV)&(MeV)&(GeV$^{-1}$)\cr
\noalign{\bigskip\hrule\bigskip}
(1,1)&780$\pm$ 3&195 $\pm$ 8&5.9 $\pm$ 1.1&1.64 $\pm$ 0.05&0.670\cr
\noalign{\bigskip\hrule\bigskip}
(1,2)&779$\pm$ 5&189 $\pm$ 12&4.3 $\pm$ 1.3&1.59 $\pm$ 0.07&0.214\cr
\noalign{\bigskip\hrule\bigskip}
(2,1)&778$\pm$ 5&192 $\pm$ 11&4.7 $\pm$ 1.3&1.59 $\pm$ 0.07&0.206\cr
\noalign{\bigskip\hrule\bigskip}
(2,2)&779$\pm$ 5&182 $\pm$ 12&3.6 $\pm$ 1.0&1.55 $\pm$ 0.07&0.212\cr
\noalign{\bigskip\hrule}\cr}
\bigskip
\bigskip
\noindent
{\bf Table 5.} Results of the fits to the four solutions of
$P$-wave intensity $I_L$ in $\pi^+ n \to \pi^+ \pi^- p$ at 5.98
GeV/c using the parametrization (4.1).
\vfill\eject
\noindent
{\bf Figure Captions}
\medskip
\item{Fig. 1.} Number of physical solutions out of 30~000 Monte
Carlo selections of SDM elements as a function of dipion mass for
$\pi^- p\to \pi^-\pi^+ n$ at 17.2 GeV/c and $-t=$0.005--0.2
(GeV/c),$^2$ and for $\pi^+ n\to \pi^+\pi^- p$ at 5.98 GeV/c and
$-t=$0.2--0.4 (GeV/c).$^2$
\item{Fig. 2} Distributions of physical values of $\cos
(\gamma_{\lower.5ex\hbox{$\scriptstyle {SL}$}})$ for $\pi^+ n\to
\pi^+\pi^- p$ at 5.98 GeV/c and $-t=$0.2--0.4 (GeV/c)$^2$ in the
dipion mass bin $520 \le m \le 600$ MeV. The distribution for
Solution 1 yields a range of
$\cos(\gamma_{\lower.5ex\hbox{$\scriptstyle {SL}$}})$ from 0.52 to
1.00 with an average value of 0.94. The distribution for Solution 2
yields a range of $\cos (\gamma_{\lower.5ex\hbox{$\scriptstyle
{SL}$}})$ from 0.31 to 1.00 with an average of 0.64.
\item{Fig. 3.} The mass dependence of physical solutions for moduli
squared of $S$-wave and $P$-wave nucleon transversity amplitudes
and cosines of their relative phases in the reaction $\pi^- p \to
\pi^- \pi^+ n$ at 17.29 GeV/c and momentum transfers $-t=$0.005--0.2
(GeV/c).$^2$ The results are in the $t$-channel dipion helicity
frame.
\item{Fig. 4.} The mass dependence of physical solutions for moduli
squared of $S$-wave and $P$-wave nucleon transversity amplitudes
and cosines of their relative phases in the reaction $\pi^+ n \to
\pi^+ \pi^- p$ at 5.98 GeV/c and momentum transfers $-t=$0.2--0.4
(GeV/c).$^2$ The results are in the $t$-channel dipion helicity
frame.
\item{Fig. 5.} The mass dependence of physical solutions for moduli
squared of $S$-wave transversity amplitudes $\overline S$ and $S$
in $\pi^- p \to \pi^-\pi^+ n$ at 17.2 GeV/c and momentum transfer
$-t =$0.005--0.2 (GeV/c).$^2$ Both solutions for the amplitude
$|\overline S|^2\Sigma$ resonate at 750--770 MeV while the amplitude
$|S|^2\Sigma$ is nonresonating in both solutions.
\item{Fig. 6.} Four solutions for $S$-wave partial wave intensity
$I_S$ in the reaction $\pi^- p \to \pi^-\pi^+ n$ at 17.2 GeV/c and
$-t=$0.005--0.2 (GeV/c).$^2$ The curves are fits to the Breit-Wigner
form (4.1). The fitted parameters are given in Table 2.
\item{Fig. 7.} Four solutions for $S$-wave partial wave intensity
$I_S$ in the reaction $\pi^+ n \to \pi^+\pi^- p$ at 5.98 GeV/c and
$-t=$0.2--0.4 (GeV/c).$^2$ The curves are fits to the Breit-Wigner
form (4.1). The fitted parameters are given in Table 3.
\item{Fig. 8.} Four solutions for $S$-wave partial wave intensity
$I_S$ in the reaction $\pi^+ n\to \pi^+ \pi^- p$ at 11.85 GeV/c and
$-t=$0.2--0.4 (GeV/c).$^2$
\item{Fig. 9.} Comparison of mass dependence of unnormalized moduli
of $P$-wave production amplitudes and associated partial wave
intensities in reaction $\pi^+ n_\uparrow \to \pi^+\pi^- p$ at 5.98
GeV/c and $-t=$0.2--0.4 (GeV/c).$^2$ Shown are solutions 1. The
other combinations of solutions are similar. The lines are to guide
the eye.
\item{Fig. 10.} Comparison of mass dependence of unnormalized
moduli of $P$-wave production amplitudes and associated partial wave
intensities in reaction $\pi^- p \to \pi^-\pi^+ n$ at 17.2 GeV/c
and $-t=$0.005--0.2 (GeV/c).$^2$ Shown are solutions 1. The other
combinations of solutions are similar. Also shown is the
Briet-Wigner fit to spin-averaged partial wave intensity $I_L(1,1)$.
The Breit-Wigner fit is scaled and compared with the moduli of spin
amplitudes $|\overline L|^2\Sigma$ and $|L|^2\Sigma$.
\item{Fig. 11.} Breit-Wigner fit to spin-averaged partial wave
intensity $I_L (1,1)$ in $\pi^+ n \to \pi^- \pi^+ p$ at 5.98 GeV/c
for $-t = 0.2 -0.4$ (GeV/c).$^2$ The Breit-Wigner fit is scaled and
compared with the moduli of spin amplitudes $|\overline L |^2\Sigma$
and $|L|^2\Sigma$.
\item{Fig. 12.} Test of predictions $\rho_{ss}^y + \rho_{00}^y +
2\rho_{11}^y = -2 (\rho_{00}^y - \rho_{11}^y) = + 2 \rho_{1-1}^y$
due to vanishing $A_1$-exchange in the reaction $\pi^- p \to
\pi^-\pi^+ n$ at 17.2 GeV/c and $-t=$0.005--0.2 (GeV/c).$^2$
\item{Fig. 13.} Test of predictions $Re \rho_{10}^y = Re
\rho_{1s}^y = Re \rho_{0s}^y = 0$ due to vanishing $A_1$-exchange in
the reaction $\pi^- p \to \pi^- \pi^+ n$ at 17.2 GeV/c and
$-t=$0.005--0.2 (GeV/c).$^2$
\item{Fig. 14.} $S$-wave intensity normalized to one at maximum
value. The data correspond to solutions $I_S(1,1)$ and $I_S(2,2)$ at
17.2 GeV/c. The smooth curves are predictions based on $\pi^+ \pi^-
\to \pi^+ \pi^-$ analyses (A and D from Ref.~27, B from Ref.~32 and
C from Ref.~33) and on current algebra and PCAC (E from Ref.~34).
The curves are taken from Ref.~15. Notice that the curves correspond
to $I=0$ $S$-wave intensity in $\pi^+\pi^- \to \pi^+ \pi^-$
reaction while the data correspond to $S$-wave intensity in $\pi^- p
\to \pi^-\pi^+ n$ in the physical region of $t$ and include
contribution from $I=2$ $S$-waves.
\item{Fig. 15.} The $I=2$ contribution to $S$-wave intensity in
$\pi^+ \pi^-$ scattering. The solid curve is $I=0$ $S$-wave
intensity in $\pi^+ \pi^- \to \pi^+ \pi^-$ from Ref.~35. The dashed
curve shows this $S$-wave intensity with $I=2$ contribution
included. The $I=2$ phase shifts were taken from Ref.~37. The data
correspond to solution $I_S (1,1)$ and $I_S(2,2)$ at 17.2 GeV/c.
\bye